\documentclass[preprintnumbers,amsmath,amssymb,floatfix,12pt,prd,superscriptaddress,nofootinbib]{revtex4}
\usepackage{graphicx}
\usepackage{epsfig}
\usepackage{bm}
\usepackage{amsfonts}
\usepackage{subfigure}

\def\bea{\begin{eqnarray}}
\def\eea{\end{eqnarray}}

\begin{document}
\begin{center}
\LARGE { \bf  Tachyon  Warm-Logamediate  Inflationary Universe
Model in High Dissipative Regime
  }
\end{center}
\begin{center}
{\bf M. R. Setare\footnote{rezakord@ipm.ir} \\  V. Kamali\footnote{vkamali1362@gmail.com}}\\
 { Department of Science, University of Kurdistan,
Sanandaj, IRAN.}
 \\
 \end{center}
\vskip 3cm

\begin{center}
{\bf{Abstract}}\\
In the present work, we study warm tachyon inflation model in the context of "logamediate inflation" where the cosmological scale factor expands as $a=a_0\exp(A[\ln t]^{\lambda})$.
The characteristics  of this model in slow-roll approximation are presented in two cases: 1- Dissipative parameter $\Gamma$ is a function of tachyon field $\phi$.
2- $\Gamma$ is a constant parameter. The level of non-Gaussianity of this model is found for these two cases. Scalar and tensor perturbations for this scenario are presented.
The parameters appearing in our model are constrained by recent observational data (WMAP7).
Harrison-Zeldovich spectrum, i.e. $n_s=1,$ is approximately obtained for large values of $\lambda$ (i.e. $\lambda\simeq 50$)and normal value of number of e-folds, i.e. $N\simeq 60$.
On the other hand, the scale invariant spectrum (Harrison-Zeldovich spectrum) is given by using two parameters in our model. For intermediate inflation,
where the cosmological scale factor expands as $a=a_0\exp(A t^{f}),$  this spectrum has  been exactly obtained using one parameter, i.e. $f=\frac{2}{3}$ \cite{2-n}.

 \end{center}
PACS numbers: 98.80.Cq~~~~~~~~~~~~~~~~~~~~~~~~~~~~~~~~~~~~~~~~~~~~~~~~~~~~~~~~~~~~~~~~~~~\\
Keywords: Warm inflation, Tachyon field, logamediate inflation, slow-roll approximation, WMAP data

\newpage

\section{Introduction}
Big Bang model has many long-standing problems (horizon,
flatness,...). These problems are solved in a framework of
inflationary universe model \cite{1-i}. Scalar field as a source
of inflation provides the causal interpretation of the origin of
the distribution of large scale structure and also observed anisotropy
of cosmological microwave background (CMB) \cite{2-i}. Standard models
for inflationary universe are divided into two regimes,
slow-roll and reheating epochs. In
slow-roll period kinetic energy remains small compared to the
potential terms. In this period, all interactions between scalar
fields (inflatons) and  other fields are neglected and as a result the
universe inflates. Subsequently, in reheating period, the kinetic
energy  is comparable to  the potential energy that causes the inflaton
starts an oscillation around  minimum of the potential losing
their energy to other fields present in the theory. After this period, the
universe is filled with radiation. \\ In warm inflation
scenario  radiation production occurs during inflationary period and
reheating is avoided \cite{4}. Thermal fluctuations may be generated
during warm inflationary era. These fluctuations could play a
dominant role to produce initial fluctuations which are necessary
for Large-Scale Structure (LSS) formation. In this model, density
fluctuation arises from thermal rather than quantum fluctuation
\cite{3-i}. Warm inflationary period ends when the universe stops
inflating. After this period the universe enters in the radiation
phase smoothly \cite{4}. Finally, remaining inflatons or dominant
radiation fields create matter components of the universe.
Some extensions of this model are found in Ref.\cite{new}.\\
Friedmann-Robertson-Walker (FRW) cosmological models in the
context of string/M-theory have related to brane-antibrane
configurations \cite{4-i}. Tachyon fields associated with
unstable D-branes may be responsible for inflation in early time
\cite{5-i}. If the tachyon field start to roll down the
potential, then universe dominated by a new form of matter will
smoothly evolve from inflationary universe to an era which is
dominated by a non-relativistic fluid \cite{1}. So, we could
explain the phase of acceleration expansion (inflation) in term
of tachyon field.\\
It has been shown that, there are  eight possible asymptotic solutions for cosmological dynamics \cite{v1}. Three of these solutions have non-inflationary scale factor and another three one's of solutions give de Sitter (with scale factor $a(t)=a_0\exp(H_0 t)$), intermediate (with scale factor $a(t)=a_0\exp(At^f), 0<f<1$) and power-low (with scale factor $a(t)=t^p, p>1$), inflationary expansions. Finally, two cases of these solutions have asymptotic expansion with scale factor($a=a_0\exp(A(\ln t)^{\lambda})$. This version of inflation is named "logamediate inflation". This model is found in a number of scalar-tensor theories \cite{1-n}. We have considered  Warm-tachyon inflationary model in the scenario of "intermedaite inflation" in Ref. \cite{2-n}. The expansion scale factor of FRW universe evolves as: $a(t)=a_0\exp(At^f)$ where $A>0$ and $ 0<f<1$. The expansion of the universe is faster than power-low inflation model and slower than de Sitter which arises as $f=1$. It was shown, to first order, the Harrizon-Zeldovich spectrum \cite{3-n} of density perturbation (i.e. $n_s=1$ \cite{4-n}) for warm-tachyon inflation arises when $f=\frac{2}{3}$.\\
In this work we would like to consider the
tachyonic warm inflationary universe in the particular scenario
"logamediate inflation" which is denoted by scale factor
$a(t)=a_0\exp(A[\ln t]^{\lambda}),\lambda>1, A>0$ \cite{v1}, \cite{6-i}.
For $\lambda=1$ case this model is converted to power-law inflation ($a=t^A; A>1$).
The study of logamediate
inflationary model is motivated by imposing weak general conditions on the cosmological model which has indefinite expansion  \cite{v1}. The effective potential of this model which arises with the above scale factor has been studied in dark energy models \cite{v2}. This potential also is found in supergravity, super-string models and Kaluza-Klein theories \cite{v3}. For this inflationary model the power spectrum may be either blue or red tilted in term of the value of the parameters of the model \cite{6-i}. Curvaton reheating in  logamediate inflationary models have been  studied in Ref.\cite{5-n}.   \\
The tachyon inflation is a k-inflation model \cite{n-1} for
scalar field $\phi$ with a positive potential $V(\phi)$. Tachyon
potentials have two special properties: 1- A maximum of
these potentials is obtained where $\phi\rightarrow 0.$ 2- A minimum of these potentials is obtained where
$\phi\rightarrow \infty$. In our  logamediate model we find an
exact solution for tachyon field potential in slow-roll
approximation which has the form $V(\phi)\sim(\ln [\phi]^{-\beta}/\phi^4)$.
This form of potential has the above properties of tachyon field
potential. This special form of potential is everlasting, so we
consider our model in the context of the warm inflation to bring
an end for inflation period. In the warm inflationary models,
dissipative effect arises from a friction term. This
effect could describe the mechanisms that scalar fields decay
into a thermal bath via its interaction with other fields that causes
the warm inflation ends when the universe heats up to become
radiation dominant. After the inflation period the universe smoothly
connected with the radiation Big Bang phase. Logamediate
inflation in standard gravity with the presence of tachyon field on the brane
has been considered in Ref.\cite{8}, and non-tachyon warm-logamediate
inflationary model has been considered in Ref.\cite{6-n}.  Also, the warm tachyon
inflation in standard gravity has been studied
 in Ref.\cite{3}. To the best of our knowledge, warm tachyon inflation model
in the context of logamediate scenario has not been yet
considered. \\The paper is organized as: In section II,
we give a brief review about the tachyonic warm inflationary universe and
its perturbation parameters in high dissipative regime.
In section III, we consider high dissipative warm-logamediate
inflationary phase in two cases: 1- Dissipative parameter $\Gamma$ as a
function of tachyon field $\phi$. 2- A constant dissipative
parameter $\Gamma$.  In this section we also,
investigate the cosmological perturbations and non-Gaussianity.
Finally in section IV, we close by concluding remarks.
\section{Tachyon warm  inflationary universe}
Tachyonic inflation model in a spatially flat Friedmann Robertson
Walker (FRW) is described by an effective fluid which is
recognized by energy-momentum tensor
$T^{\mu}_{\nu}=diag(-\rho_{\phi},p_{\phi},p_{\phi},p_{\phi})$
\cite{1}, where energy density $\rho_{\phi}$, and pressure for
the tachyon field are defined by
\begin{eqnarray}\label{1}
\rho_{\phi}=\frac{V(\phi)}{\sqrt{1-\dot{\phi}^2}}~~~~~~~~\\
\nonumber p_{\phi}=-V(\phi)\sqrt{1-\dot{\phi}^2},
\end{eqnarray}
respectively, where $\phi$ denotes the tachyon scalar field and
$V(\phi)$ is the effective scalar potential associated with the
tachyon field. Characteristics of any tachyon field potential are
$\frac{dV}{d\phi}<0$ and $V(\phi\rightarrow 0)\rightarrow V_{max}$
\cite{2}. Friedmann equation for spatially flat universe and
conservation equation, in the warm tachyon inflationary scenario
are given by \cite{3}
\begin{eqnarray}\label{2}
 H^2=\frac{8\pi}{3m_p^2}[\rho_{\phi}+\rho_{\gamma}],
\end{eqnarray}

\begin{eqnarray}\label{3}
\dot{\rho}_{\phi}+3H(\rho_{\phi}+p_{\phi})=-\Gamma
\dot{\phi}^2\Rightarrow~~~~\frac{\ddot{\phi}}{1-\dot{\phi}^2}+3H\dot{\phi}+\frac{V'}{V}=-\frac{\Gamma}{V}\sqrt{1-\dot{\phi}^2}\dot{\phi},
\end{eqnarray}
and
\begin{eqnarray}\label{4}
 \dot{\rho}_{\gamma}+4H\rho_{\gamma}=\Gamma \dot{\phi}^2,
\end{eqnarray}
where $H=\frac{\dot{a}}{a},$ is Hubble parameter, $a$ is scale
factor, $m_p$ represents the Planck mass and  $\rho_{\gamma}$ is
energy density of the radiation. Dissipative coefficient
$\Gamma$ has the dimension $m_p^5$. In the above equations dots
$"^{.}"$ mean derivative with respect to time and prime $"'"$ is
derivative with respect to $\phi$. During inflation epoch
the energy density (\ref{1}) is the order of the potential, i.e.
$\rho_{\phi}\sim V$. Tachyon energy density $\rho_{\phi}$ in this
era dominates over the energy density of radiation, i.e.
$\rho_{\phi}>\rho_{\gamma}$. In slow-roll regime, i.e. $\dot{\phi}\ll
1$ and $\ddot{\phi}\ll(3H+\Gamma/V)\dot{\phi}$ \cite{4} and when
the radiation production in warm inflation era is quasi-stable, i.e. $\dot{\rho}_{\gamma}\ll 4H\rho_{\gamma},
\dot{\rho}_{\gamma}\ll\Gamma\dot{\phi}^2,$ the equations
(\ref{2}), (\ref{3}) and (\ref{4}) are reduced to
\begin{eqnarray}\label{5}
H^2=\frac{8\pi}{3m_p^2}V,
\end{eqnarray}

\begin{eqnarray}\label{6}
3H(1+r)\dot{\phi}=-\frac{V'}{V},
\end{eqnarray}
\begin{eqnarray}\label{7}
\rho_{\gamma}=\frac{\Gamma\dot{\phi}^2}{4H},
  \end{eqnarray}
where $r=\frac{\Gamma}{3HV}$. The main problem of inflation
theory is how to attach the universe to the end of the inflation
period. One of the solutions of this problem is the study of
inflation in the context of warm inflation \cite{m1}. In this
model radiation is produced during inflation period where its
energy density is kept nearly constant. This is phenomenologically
fulfilled by introducing the dissipation term $\Gamma$, in the
equation of motion as we have seen in Eq.(\ref{3}). This term
grants a  continuous energy  transfer from scalar field energy
into a thermal bath. In this article we consider high dissipation
regimen, i.e. $r\gg 1,$ where the dissipation coefficient $\Gamma$ is
much greater than the $3HV$. High dissipative and weak
dissipative regimes for non-tachyonic warm inflation have been
studied in Refs.\cite{m1} and \cite{m2} respectively. The
main attention was given to the high dissipative regime. This
regime is the more difficult of the two, and when high
dissipative regime is understood, the weak dissipative regime
could follow. Also, in Refs.\cite{m2} and \cite{m3}, the study of
warm inflation in the context of quantum field theory has been
fulfilled in high dissipative regime. Dissipative parameter
$\Gamma$ could be a constant or a positive function of $\phi$ by
the second law of thermodynamics. In some works $\Gamma$ and
potential of inflation have the same form\cite{3}, \cite{m4}. In
Ref.\cite{3}, perturbation parameters for warm tachyon inflation
have been obtained where $\Gamma=\Gamma_0=const$ and
$\Gamma=\Gamma(\phi)=V(\phi)$. So, in this work we will study the
logamediate tachyon warm inflation in high dissipative regime
for these two cases.
\\
From Eqs. (\ref{6}) and (\ref{7}) in high dissipation regime we
get
\begin{eqnarray}\label{8}
\rho_{\gamma}=\frac{m_p^2}{32\pi r}[\frac{V'}{V}]^2.
\end{eqnarray}
We introduce the slow-roll parameter
\begin{eqnarray}\label{9}
\epsilon=-\frac{\dot{H}}{H^2}=\frac{m_p^2}{16\pi
r}[\frac{V'}{V}]^2\frac{1}{V}.
\end{eqnarray}
Using Eqs. (\ref{8}) and (\ref{9}) in slow-roll regime
($\rho_{\phi}\sim V$) we find a relation between the energy
densities $\rho_{\phi}$ and $\rho_{\gamma}$ as
\begin{eqnarray}\label{10}
\rho_{\gamma}=\frac{\epsilon}{2}\rho_{\phi}.
\end{eqnarray}
The second slow-roll parameter $\eta$ is given by
\begin{eqnarray}\label{11}
\eta=-\frac{\ddot{H}}{H\dot{H}}\simeq\frac{m_p^2}{8r\pi V}[\frac{V''}{V}-\frac{1}{2}(\frac{V'}{V})^2].
\end{eqnarray}
The warm inflationary condition, i.e. $\ddot{a}>0,$ may be obtained by
parameter $\epsilon$, satisfying the relation
\begin{eqnarray}\label{12}
\epsilon <1.
\end{eqnarray}
From above equation and Eq.(\ref{10}) the tachyon warm inflation
epoch could take place when
\begin{equation}\label{13}
\rho_{\phi}>2\rho_{\gamma}.
\end{equation}
For the tachyon
field in warm inflationary universe (in slow-roll and high
dissipative regime) the power spectrum of the curvature
perturbation and amplitude of tensor perturbation (which would
produce gravitational waves during inflation) are given by
\cite{3}
\begin{eqnarray}\label{14}
P_R\simeq\frac{\sqrt{3}}{30\pi^2}\exp(-2\Im(\phi))[(\frac{1}{\epsilon})^3\frac{9m_p^4}{128\pi^2r^2\sigma
V}]^{\frac{1}{4}}
\end{eqnarray}
\begin{eqnarray}\label{15}
P_T=\frac{16\pi}{m_p^2}(\frac{H}{2\pi})^2\coth[\frac{k}{2T}]\simeq\frac{32V}{3m_p^4}\coth[\frac{k}{2T}],
\end{eqnarray}
respectively. Temperature $T$ in extra factor
$\coth[\frac{k}{2T}]$ denotes the temperature of the thermal
background of gravitational wave \cite{6} and
\begin{eqnarray}\label{16}
\Im(\phi)=-\int[\frac{1}{3Hr}(\frac{\Gamma}{V})'+\frac{9}{8}\frac{V'}{V}[1-\frac{(\ln
\Gamma)'(\ln V)'}{36H^2r}]]d\phi.
\end{eqnarray}
In high dissipative regime ($r\gg 1$), from Eqs.(\ref{14}) and (\ref{15}), tensor-scalar
ratio  is obtained
\begin{eqnarray}\label{17}
R(k_0)\approx
\frac{240\sqrt{3}}{25m_p^2}[\frac{r^{\frac{1}{2}}\epsilon
H^3}{T_r}\exp[2\Im(\phi)]\coth(\frac{k}{2T})]|_{k=k_0}.
\end{eqnarray}
$R$ is important parameter. We can use the seven-year  Wilkinson
Microwave Anisotropy Probe (WMAP7) data to find an upper
bound for $R$, from these results  we have $P_R\simeq 2.28\times
10^{-9}, R=0.21<0.36$  \cite{2-i}. Spectral indices $n_g$ and
$n_s$ were presented in \cite{3}
\begin{eqnarray}\label{18}
n_g=-2\epsilon~~~~~~~~~~~~~~~~~~~~~~~~~~~~~~~~~~~~~~~~,\\
\nonumber
n_s=1-[\frac{3}{2}\eta+\epsilon(\frac{2V}{V'}[2\Im'(\phi)-\frac{r'}{4r}]-\frac{5}{2})].
\end{eqnarray}
From above equations we could find interesting inflationary parameters in slow-roll approximation in the context of logamediate inflation. The non-Gaussainty level for warm tachyon inflation in high dissipative regime may be obtained by $f_{NL}$ parameter ―cite{2-n}
\begin{eqnarray}\label{Ga}
f_{NL}=-\frac{5}{3}(\frac{\dot{\phi}}{H})[\frac{1}{H}\ln(\frac{k_F}{H})(\frac{V'''(\phi_0(t_F))+2k_F^2V'(\phi_0(t_F))}{\Gamma})]
\end{eqnarray}
where $t_F$ is the time when the last three wave-vector thermalize and $k_{F}=\sqrt{\frac{\Gamma H}{V}}$ is freeze-out momentum.
\section{Logamediate inflation}
In this section we consider high dissipative warm tachyon logamediate
inflation. In logamediate inflation the scale factor follows the
law
\begin{eqnarray}\label{19}
a(t)=a_0\exp(A[\ln t]^{\lambda}) ~~~~~~\lambda>1,
\end{eqnarray}
where $A$ is a positive constant. From above scale factor, the number of e-folds may be found
\begin{eqnarray}\label{20}
N=\int_{t1}^{t2} H dt=A[(\ln t_2)^{\lambda}-(\ln t_1)^{\lambda}]
\end{eqnarray}
We would like to consider our model in two important cases \cite{3}: 1- $\Gamma$ is a function of
$\phi$ ($\Gamma=f(\phi)=\Gamma_1V(\phi)$). 2- $\Gamma$ is a constant parameter.
\subsection{$\Gamma=\Gamma{(\phi)}=\Gamma_1V(\phi)$}
At the late time and by using Eqs. (\ref{5}), (\ref{6}) and (\ref{19}) with
$\Gamma=\Gamma_1V(\phi)$ where $\Gamma_1>0$, we get tachyon scalar field $\phi$
\begin{eqnarray}\label{21}
\Gamma_1\dot{\phi}^2=-\frac{2\dot{H}}{H}=\frac{2}{t}\Rightarrow \sqrt{\Gamma_1}(\phi-\phi_0)=2\sqrt{2t}
\end{eqnarray}
where
\begin{eqnarray}\label{22}
\dot{H}=\frac{\lambda A(\ln t)^{\lambda-1}}{t^2}[\frac{\lambda-1}{\ln t}-1]
\end{eqnarray}
At the late time the first term in the above equation is omitted.
Effective potential in terms of tachyon field is given by
\begin{eqnarray}\label{23}
V(\phi)=\frac{24 m_P^2(\lambda A)^2}{\Gamma_1^2\pi}\frac{[\ln(\Gamma_1\frac{(\phi-\phi_0)^2}{8})]^{2\lambda-2}}{(\phi-\phi_0)^4}
\end{eqnarray}
The Hubble parameter in term of tachyon field $\phi$ becomes
\begin{eqnarray}\label{24}
H(\phi)=\frac{\lambda A [\ln(\Gamma_1\frac{(\phi-\phi_0)^2}{8})]^{\lambda-1}}{\Gamma_1(\phi-\phi_0)^2}
\end{eqnarray}
From equation (\ref{23}),  $V(\phi)$ has the
characteristic of tachyon field potential ($\frac{dV}{d\phi}<0$
and $V(\phi\rightarrow 0)\rightarrow V_{max}$), also these potentials at the late time
do not have a minimum \cite{5}. Slow-roll parameters $\epsilon$
and $\eta$ in term of tachyon field $\phi$ are obtained from Eqs.
(\ref{9}) and (\ref{11}).
\begin{eqnarray}\label{25}
\epsilon =―frac{[\ln(\Gamma_1\frac{(\phi-\phi_0)^2}{8})]^{1-\lambda}}{\lambda A}
\end{eqnarray}

\begin{eqnarray}\label{26}
\eta =\frac{2[\ln(\Gamma_1\frac{(\phi-\phi_0)^2}{8})]^{1-\lambda}}{\lambda A}
\end{eqnarray}
respectively. From Eq.(\ref{20}), the number of e-folds between
two fields $\phi_1=\phi(t_1)$ and $\phi_2=\phi(t_2)$
 is given by
\begin{eqnarray}\label{27}
N(\phi)=A([\ln(\Gamma_1\frac{(\phi_2-\phi_0)^2}{8})]^{\lambda}-[\ln(\Gamma_1\frac{(\phi_1-\phi_0)^2}{8})]^{\lambda})
\end{eqnarray}
$\phi_1$ is obtained at the begining of inflation ($\epsilon=1$),
$\phi_1=\frac{2\sqrt{2}}{\sqrt{\Gamma_1}}\exp(\frac{(\lambda A)^{\frac{1}{1-\lambda}}}{2})+\phi_0$.
The value of $\phi_2$  could be determined in terms of $N$, $A$ and $\lambda$ parameters
\begin{eqnarray}\label{28}
\phi_2=\frac{2\sqrt{2}}{\sqrt{\Gamma_1}}\exp([\frac{1}{2}(\frac{N}{A}+(\lambda A)^{\frac{\lambda}{1-\lambda}})^{\frac{1}{\lambda}}])+\phi_0
\end{eqnarray}

In logamediate inflation, we obtain perturbation  parameters in term
of tachyon field. Firstly, from Eq.(\ref{16}) in this case ($\Gamma=\Gamma_1V$) we find
\begin{eqnarray}\label{29}
\Im(\phi)_1=-―frac{9}{8}\ln[V(\phi)]-\frac{3―Gamma_1}{8\lambda A}\gamma[2-\lambda,\ln(\Gamma_1\frac{(\phi-\phi_0)^2}{8})]
\end{eqnarray}
where $\gamma[a,t],$ is incomplete gamma function \cite{gamma}.
From above equation and Eq.(\ref{14}) we obtain spectrum of curvature perturbation in slow-roll limit
\begin{eqnarray}\label{}
P_R=\alpha [\ln(\Gamma_1\frac{(\phi-\phi_0)^2}{8})]^{\frac{3}{4}(\lambda-1)}
\end{eqnarray}
where $\alpha=\frac{\exp(-2\Im(\phi)_1)}{10\Gamma_1\pi^2}(\frac{3m_p^2\Gamma_1^2(\lambda A)^3}{1024\pi\sigma})^{\frac{1}{4}}$. In high dissipative regime ($r\gg 1$) we have
\begin{eqnarray}\label{30}
P_R=\beta\frac{[\ln(\Gamma_1\frac{(\phi-\phi_0)^2}{8})]^{\frac{21}{4}(\lambda-1)}}{(\phi-\phi_0)^4}~~~~~~~~~~~~~~~~~~~~~~~~~~~~~~~~~~~~~\\
\nonumber
=\frac{\Gamma_1^2\beta}{64}\exp(-2(\frac{N}{A}+(\lambda A)^{\frac{\lambda}{1-\lambda}})^{\frac{1}{\lambda}})(\frac{N}{A}+(\lambda A)^{\frac{\lambda}{1-\lambda}})^{\frac{21(\lambda-1)}{4\lambda}}
\end{eqnarray}
where $\beta=\frac{1}{10\Gamma_1\pi^2}(\frac{3m_p^2\Gamma_1^2(\lambda A)^3}{1024\pi\sigma})^{\frac{1}{4}}(\frac{24m_p^2(\lambda A)^2}{\Gamma_1^2\pi})^{\frac{9}{4}}$.
This parameter may be constrained by WMAP7 data \cite{2-i}.
The amplitude of tensor perturbation in this case, from Eq.(\ref{15}) becomes
\begin{eqnarray}\label{31}
P_T=\frac{256 (\lambda A)^2}{\Gamma_1^2\pi}\frac{[\ln(\Gamma_1\frac{(\phi-\phi_0)^2}{8})]^{2\lambda-2}}{(\phi-\phi_0)^4}\coth[\frac{k}{2T}]~~~~~~~~~~~~~~~~~~~~~~~~~~~~~\\
\nonumber
=\frac{4(\lambda A)^2}{\pi}\exp(-2(\frac{N}{A}+(\lambda A)^{\frac{\lambda}{1-\lambda}})^{\frac{1}{\lambda}})(\frac{N}{A}+(\lambda A)^{\frac{\lambda}{1-\lambda}})^{\frac{2\lambda-1}{\lambda}}\coth[\frac{k}{2T}]
\end{eqnarray}
From Eq.(\ref{18}) the spectral indices $n_g$ and $n_s$  are given by
\begin{eqnarray}\label{32}
n_g=-2\frac{[\ln(\Gamma_1\frac{(\phi-\phi_0)^2}{8})]^{1-\lambda}}{\lambda A}~~~~~~~~~~~~~~~~~~~~~~~~~~~~~~~~\\
\nonumber
n_s=1-\frac{3}{2}\eta+\frac{3}{4}\epsilon(1+\epsilon)\simeq 1-\frac{9[\ln(\Gamma_1\frac{(\phi-\phi_0)^2}{8})]^{1-\lambda}}{4\lambda A}\\
\nonumber
=1-\frac{9}{4\lambda A}(\frac{N}{A}+(\lambda A)^{\frac{\lambda}{1-\lambda}})^{\frac{1-\lambda}{\lambda}}
\end{eqnarray}
In Fig.(1), the dependence of spectral index on the number of e-folds of inflation is shown (for $\lambda=5$ and $\lambda=50$ cases). It is  observed that small values of number of e-folds are assured for large values of $\lambda$ parameter.
\begin{figure}[h]
\begin{minipage}[b]{1\textwidth}
\subfigure[\label{fig1a} ]{ \includegraphics[width=.37\textwidth]%
{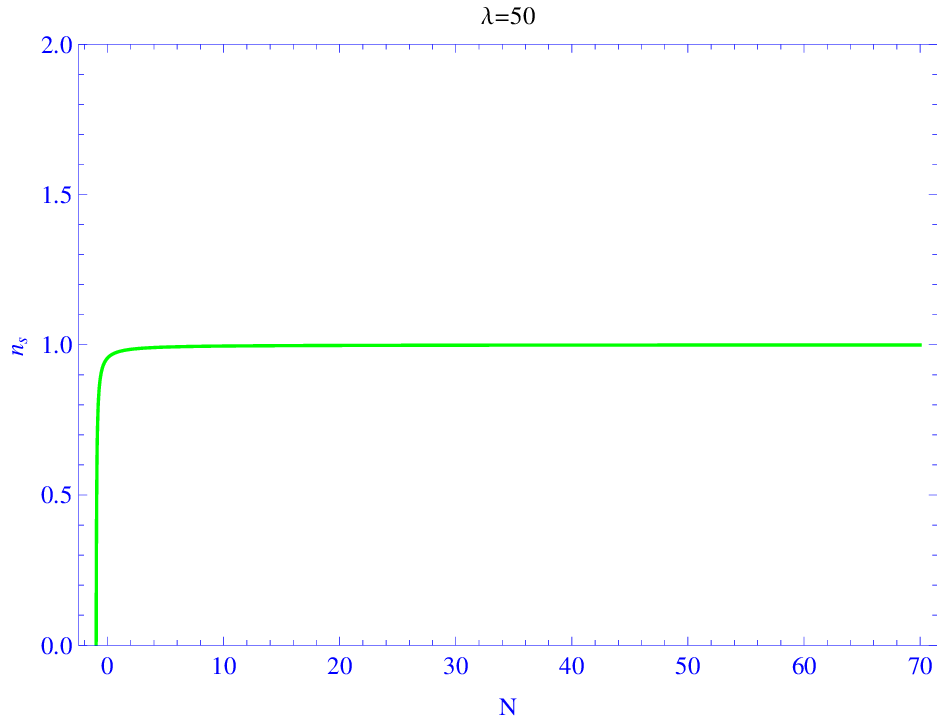}} \hspace{.2cm}
\subfigure[\label{fig1b} ]{ \includegraphics[width=.37\textwidth]%
{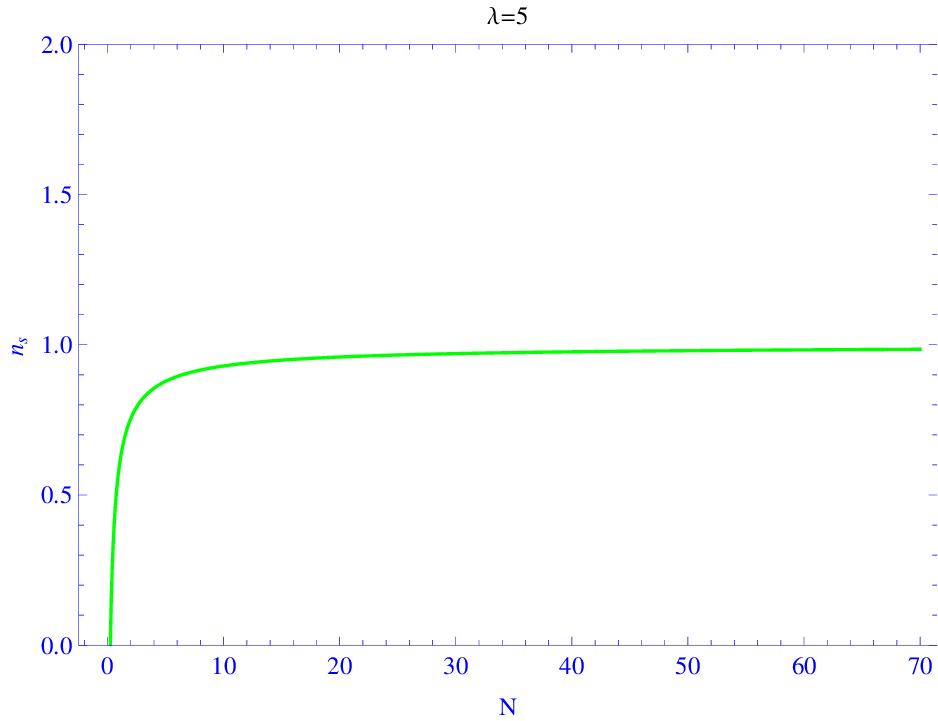}}
\end{minipage}
\caption{ Spectral index $n_s$ in term of number of e-folds: (a) for $\lambda=50$ and (b) for $\lambda=5$ (where $A=1$).}
\end{figure}
From Eq.(\ref{17}), we could find the tensor-scalar ratio  as
\begin{eqnarray}\label{33}
R=\frac{256(\lambda A)^2}{\Gamma_1^2\beta}(\frac{N}{A}+(\lambda A)^{\frac{\lambda}{1-\lambda}})^{\frac{-13\lambda+17}{4\lambda}}\coth[\frac{k}{2T}]
\end{eqnarray}
The above  parameter is found from WMAP7 observational data.
Non-Gaussianity in term of tachyon field in this case is obtained from Eq.(\ref{Ga})
 \begin{eqnarray}\label{34}
f_{NL}=\frac{5}{3}\Gamma_1\ln(\Gamma_1)[\ln(\frac{\Gamma_1(\phi-\phi_0)^2}{8})]^{1-\lambda}=\frac{5}{3}\Gamma_1\ln(\Gamma_1)[\frac{N}{A}+(\lambda A)^{\frac{1}{1-\lambda}}]^{\frac{1-\lambda}{\lambda}}
\end{eqnarray}
Where $\lambda>1$ this parameter have small amount at the late time. FIG.(1) shows  the Harrison-Zeldovich spectrum, i.e. $n_s=1,$ could approximately obtained for $(\lambda,N)=(50,60)$. In this case, we have small level of non-Gaussianity.
From Eq.(\ref{32}) and (\ref{33}) we can find the tensor-scalar ratio $R$ versus spectral index $n_s$.
 \begin{eqnarray}\label{}
R=\frac{256(\lambda A)^2}{\Gamma_1^2\beta}(\frac{4\lambda A}{9}(1-n_s))^{\frac{13\lambda-17}{4(\lambda-1)}}\coth[\frac{k}{2T}]
\end{eqnarray}
In Fig.(2), two trajectories in the $n_s-R$ plane are shown. There is a range of values of $R$ and $n_s$ which is compatible with the WMAP7 data. The scale-invariant spectrum (Harrison-Zeldovich spectrum, i.e. $n_s=1$)  may be obtained for $(\lambda,\Gamma_1)=(50,37.5)$.
\begin{figure}[h]
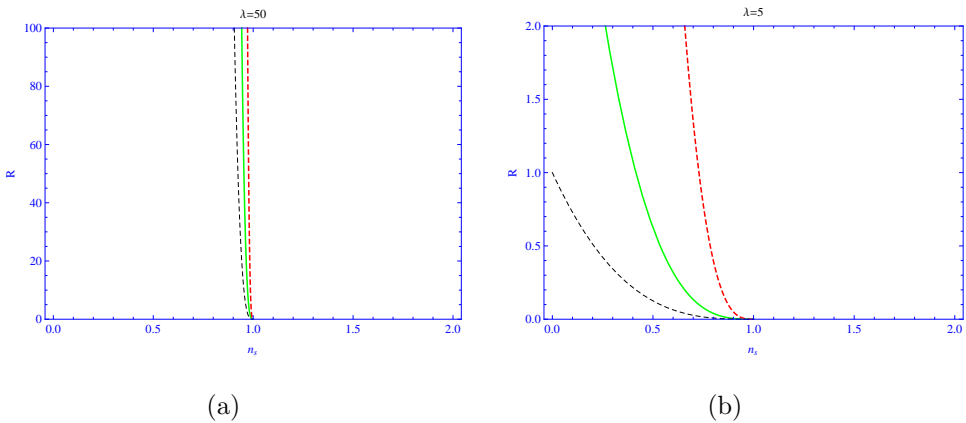

\begin{minipage}[b]{1\textwidth}
\subfigure[\label{fig1a} ]{ \includegraphics[width=.37\textwidth]%
{g.eps}} \hspace{.2cm}
\subfigure[\label{fig1b} ]{ \includegraphics[width=.37\textwidth]%
{f.eps}}
\end{minipage}
\caption{ Tensor-scalar ratio  in term of spectral index $n_s$: (a) for $\lambda=50$ and (b) for $\lambda=5$ (where  $k_0=0.002 Mpc^{-1}$, $T=T_r=5.47\times 10^{-5}$, $A,\beta=1$, $\Gamma_1=118.5$ case with the  green line, $\Gamma_1=37.5$ case with the red dashed line, $\Gamma_1=265$ case with the  black dashed line ).}
\end{figure}

\subsection{ $\Gamma=\Gamma_0=const$}
By using Eqs. (\ref{5}), (\ref{6}) and (\ref{19}) with
$\Gamma=\Gamma_0$ at the late time , we get tachyon scalar field $\phi$
\begin{eqnarray}\label{34}
\phi-\phi_0=\frac{\Upsilon_{\lambda}[t]}{\varpi}
\end{eqnarray}
where $\varpi=\frac{\sqrt{4\pi\Gamma_0}}{m_p\sqrt{3}\lambda A 2^{\lambda}}$ and $\Upsilon=\gamma[\lambda,\frac{\ln t}{2}]$ ($\gamma[a,t]$ is incomplete gamma function \cite{gamma}). The Hubble parameter and effective potential are obtained as
\begin{eqnarray}\label{35}
H(\phi)=\frac{\lambda A[\ln\Upsilon^{-1}[\varpi(\phi-\phi_0)]]^{\lambda-1}}{\Upsilon^{-1}[\varpi(\phi-\phi_0)]}
\end{eqnarray}
and
\begin{eqnarray}\label{36}
V(\phi)=\frac{3m_p^2(\lambda A)^2[\ln\Upsilon^{-1}[\varpi(\phi-\phi_0)]]^{2\lambda-2}}{8\pi(\Upsilon^{-1}[\varpi(\phi-\phi_0)])^2}
\end{eqnarray}
respectively. $\Upsilon^{-1}$ is inverse function of the $\Upsilon$. From Eqs.(\ref{9}) and (\ref{11}) we could find slow-roll parameters $\epsilon$ and $\eta$ in term of tachyon field (\ref{34}),
\begin{eqnarray}\label{37}
\epsilon=\frac{(\ln\Upsilon^{-1}[\varpi(\phi-\phi_0)])^{1-\lambda}}{\lambda A}
\end{eqnarray}
and
\begin{eqnarray}\label{38}
\eta=\frac{2(\ln\Upsilon^{-1}[\varpi(\phi-\phi_0)])^{1-\lambda}}{\lambda A}
\end{eqnarray}
respectively. Using Eq.(\ref{20}) the number of e-folds is obtained as
\begin{eqnarray}\label{39}
N=A[(\ln[\Upsilon^{-1}_{\lambda}(\varpi(\phi_2-\phi_0)])^{\lambda}-(\ln[\Upsilon^{-1}_{\lambda}(\varpi(\phi_1-\phi_0))])^{\lambda}]
\end{eqnarray}
At the begining of inflation ($\epsilon=1$), $\phi_1=\frac{\Upsilon[\exp([A\lambda]^{\frac{1}{1-\lambda}})]}{\varpi}$, so tachyon field $\phi_2$ at the end of inflation in terms of number of e-folds becomes
\begin{eqnarray}\label{40}
\phi_2-\phi_0=\frac{\Upsilon[\exp([\frac{N}{A}+(\lambda A)^{\frac{1}{1-\lambda}}]^{\frac{1}{\lambda}})]}{\varpi}
\end{eqnarray}
In this case ($\Gamma=\Gamma_0$), perturbation parameters may be obtained in terms of tachyon field. Therefore from Eq.(\ref{16}) we have
\begin{eqnarray}\label{41}
\Im(\phi)_2=-\frac{\ln V}{8}
\end{eqnarray}
We obtain spectrum of curvature perturbation in slow-roll limit, from above equation and Eq.(\ref{14})
\begin{eqnarray}\label{42}
P_R=\alpha' \frac{(\ln\Upsilon^{-1}[\varpi(\phi-\phi_0)])^{\frac{9\lambda-9}{4}}}{(\Upsilon^{-1}[\varpi(\phi-\phi_0)])^{\frac{3}{2}}}=~~~~~~~~~~~~~~~~~~~~~~~~~~~~~\\
\nonumber
\alpha'[\frac{N}{A}+(\lambda A)^{\frac{1}{1-\lambda}}]^{\frac{9\lambda-9}{4\lambda}}\exp(-\frac{3}{2}[\frac{N}{A}+(\lambda A)^{\frac{1}{1-\lambda}}]^{\frac{1}{\lambda}})
\end{eqnarray}
where $\alpha'=\frac{\sqrt{3}}{30\pi^2}(\frac{3m_p(\lambda A)^3}{8\pi})^{\frac{3}{4}}(\frac{9}{16\pi\sigma\Gamma_0})^{\frac{1}{4}}$.
This parameter may be constrained by WMAP7 observational data \cite{2-i}.
From Eqs.(\ref{15}) and (\ref{36}), the amplitude of tensor perturbation in this case,  becomes
\begin{eqnarray}\label{43}
P_T=\frac{4(\lambda A)^2}{\pi m_p^2}\frac{(\ln\Upsilon^{-1}[\varpi(\phi-\phi_0)])^{2\lambda-2}}{(\Upsilon^{-1}[\varpi(\phi-\phi_0)])^2}=~~~~~~~~~~~~~~~~~~~~~~~~~~~~~~~~~~\\
\nonumber
\frac{4(\lambda A)^2}{\pi m_p^2}[\frac{N}{A}+(\lambda A)^{\frac{1}{1-\lambda}}]^{\frac{2\lambda-2}{\lambda}}\exp(-2[\frac{N}{A}+(\lambda A)^{\frac{1}{1-\lambda}}]^{\frac{1}{\lambda}})\coth[\frac{k}{2T}]
\end{eqnarray}
From Eq.(\ref{18}) the spectral indices $n_s$ and $n_g$ are given by
\begin{eqnarray}\label{44}
n_g=-\frac{2(\ln \Upsilon^{-1}[\varpi(\phi-\phi_0)])^{1-\lambda}}{\lambda A}~~~~~~~~~~~~~~~~~~~~~~~~~\\
\nonumber
n_s=1-\frac{3}{4}\eta-\frac{9}{4}\epsilon=1-\frac{13(\ln\Upsilon^{-1}[\varpi(\phi-\phi_0)])^{1-\lambda}}{4\lambda A}\\
\nonumber
=1-\frac{13}{4\lambda A}[\frac{N}{A}+(\lambda A)^{\frac{1}{1-\lambda}}]^{\frac{1-\lambda}{\lambda}}~~~~~~~~~~~~~~~~~
\end{eqnarray}
In Fig.(3), the dependence of spectral index on the number of e-folds of inflation is shown (for $\lambda=5$ and $\lambda=50$ cases). It is observed that small values of number of e-folds are assured for large values of $\lambda$ parameter.
\begin{figure}[h]
\begin{minipage}[b]{1\textwidth}
\subfigure[\label{fig1a} ]{ \includegraphics[width=.37\textwidth]%
{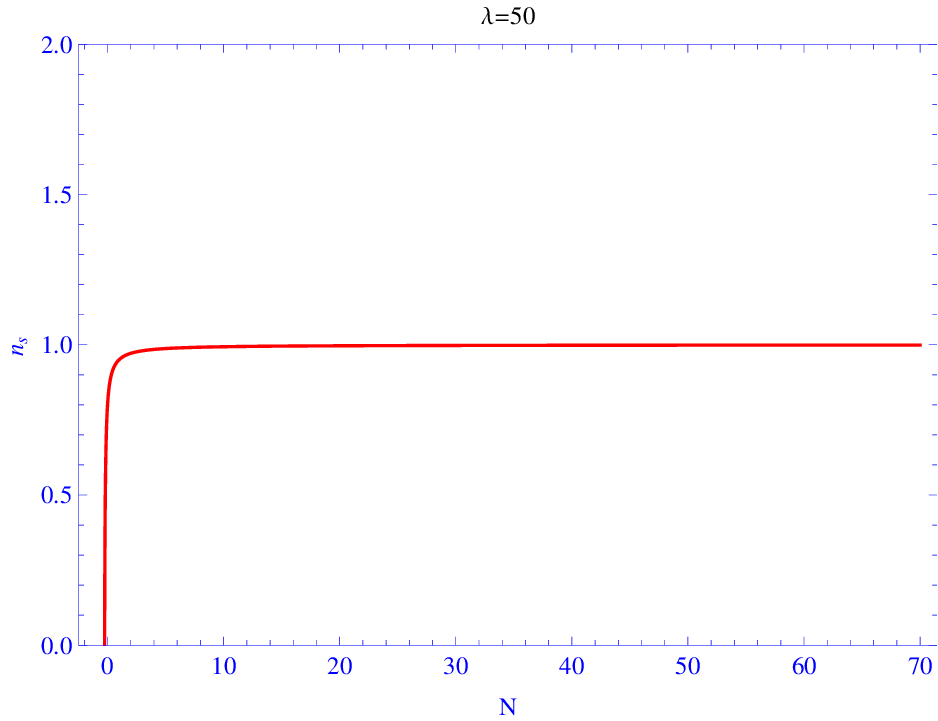}} \hspace{.2cm}
\subfigure[\label{fig1b} ]{ \includegraphics[width=.37\textwidth]%
{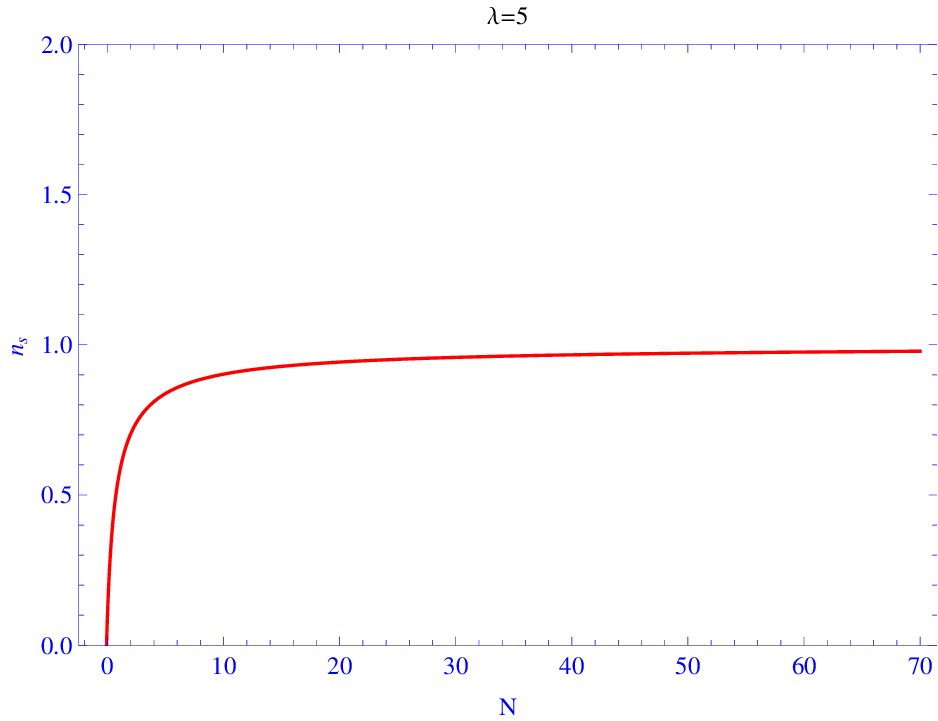}}
\end{minipage}
\caption{ Spectral index $n_s$ in term of number of e-folds: (a) for $\lambda=50$ and (b) for $\lambda=5$ (where $A=1$).}
\end{figure}
We could find the tensor-scalar ratio  as (from Eq.(\ref{17}))
\begin{eqnarray}\label{45}
R=\frac{9(\lambda A)^2}{\pi m_p^2\alpha'}[\frac{N}{A}+(\lambda A)^{\frac{1}{1-\lambda}}]^{\frac{1-\lambda}{4\lambda}}\exp(-\frac{1}{2}[\frac{N}{A}+(\lambda A)^{\frac{1}{1-\lambda}}]^{\frac{1}{\lambda}})\coth[\frac{k}{2T}]
\end{eqnarray}
this parameter is found from WMAP7 observational data \cite{2-i}. From Eq.(\ref{Ga}) the non-Gaussianity in this case is obtained as
\begin{eqnarray}\label{}
f_{NL}=\frac{5}{3}(\ln\Upsilon^{-1}[\varpi(\phi-\phi_0)])^{1-\lambda}=\frac{5}{3}[\frac{N}{A}+(\lambda A)^{\frac{1}{1-\lambda}}]^{\frac{1-\lambda}{\lambda}}
\end{eqnarray}
Where  $\lambda>1$ the non-Gaussianity has small level at the late time.
 FIG.(3) shows the scale invariant spectrum, (Harrison-Zeldovich spectrum, i.e. $n_s=1$) could be approximately obtained for ($\lambda,$$N$)=(50,60). In this case, the small level of non-Gaussianity is found.
From Eq.(\ref{44}) and (\ref{45}) we can find the tensor-scalar ratio $R$ versus spectral index $n_s$.
\begin{eqnarray}\label{}
R=\frac{9(\lambda A)^2}{\pi m_p^2\alpha'}[\frac{4\lambda A}{3}(1-n_s)]^{\frac{1}{4}}\exp(-\frac{1}{2}[\frac{4\lambda A}{3}(1-n_s)]^{\frac{1}{1-\lambda}})\coth[\frac{k}{2T}]
\end{eqnarray}
In Fig.(4), two trajectories in the $n_s-R$ plane are shown. There is a range of values of $R$ and $n_s$ which is compatible with the WMAP7 data. The scale-invariant spectrum (Harrison-Zeldovich spectrum, i.e. $n_s=1$) may be obtained for $(\lambda,\Gamma_0)=(50,37.5)$.
\begin{figure}[h]
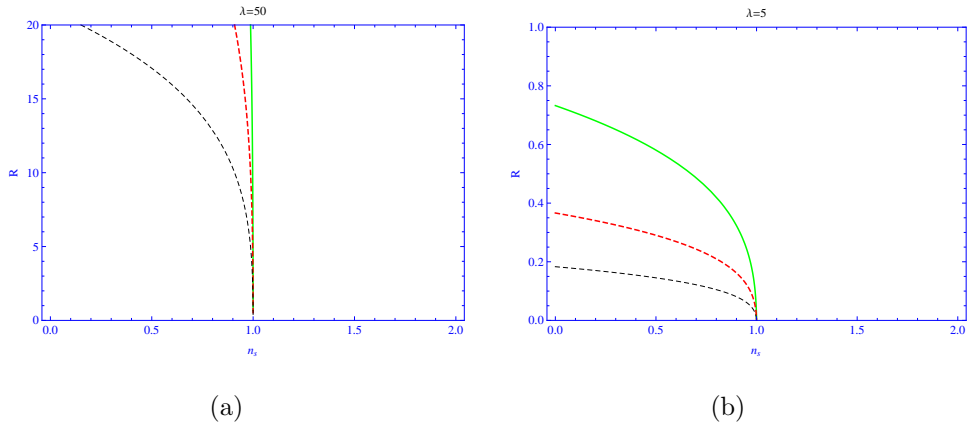

\begin{minipage}[b]{1\textwidth}
\subfigure[\label{fig1a} ]{ \includegraphics[width=.37\textwidth]%
{k.eps}} \hspace{.2cm}
\subfigure[\label{fig1b} ]{ \includegraphics[width=.37\textwidth]%
{h.eps}}
\end{minipage}
\caption{ Tensor-scalar ratio  in term of spectral index $n_s$: (a) for $\lambda=50$ and (b) for $\lambda=5$ (where   $k_0=0.002 Mpc^{-1}$, $T=T_r=5.47\times 10^{-5}$, $A,\alpha'=1$, $\Gamma_0=118.5$ case with the  green line, $\Gamma_0=37.5$ case with the red dashed line, $\Gamma_0=265$ case with the  black dashed line ).}
\end{figure}
Using WMAP7 data, $P_R(k_0)\simeq 2.28\times 10^{-9}$, $R(k_0)\simeq 0.21$ and the characteristic of warm inflation $T>H$ \cite{3},
we may restrict the values of temperature to $T_r>5.47\times 10^{-5}$ using Eqs.(\ref{14}),(\ref{17}) or equivalently Eqs.(\ref{42}), (\ref{45}), (see Fig.(5)).
We have chosen $k_0=0.002 Mpc^{-1}$ and $T\simeq T_r$.
\begin{figure}[h]
\centering
  \includegraphics[width=10cm]{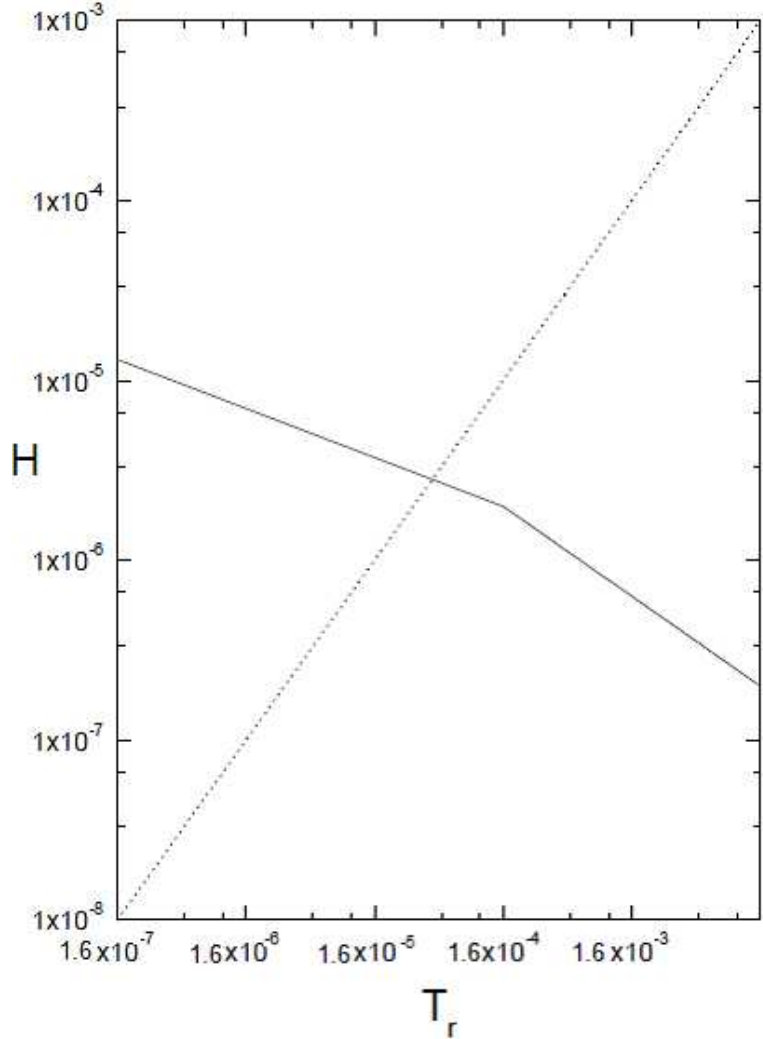}
  \caption{In this graph we plot the Hubble parameter $H$ in term of the temperature $T_r$. We can find the minimum amount of temperature $T_r=5.47\times 10^{-5}$ in order to have the necessary condition for warm inflation model ($T_r>H$). }
 \label{fig:F3}
\end{figure}

\section{Conclusion}
In this article we have investigated  the warm-tachyon-logamediate
inflationary model. In this model we have found an everlasting form of potential. This form of potential agrees with tachyon potential properties. Warm inflation is an important model as a mechanism which gives an end for inflation epoch. Therefore we have considered  the tachyon warm-logamediate inflationary model. This model has been developed for $\Gamma$ as a function of tachyon field $\phi$ and for $\Gamma=\Gamma_0$. For these two cases we have extracted the form of potential and Hubble parameters as a function of tachyon field $\phi$. Explicit expressions for tensor-scalar ratio $R$, and spectrum indices $n_g$ and $n_s$ in slow-roll were obtained. We also have constrained these parameters by WMAP7 results.\\ In $\Gamma=\Gamma(\phi)=V(\phi)$ case we have obtained the form of tachyon field as $\phi\propto\sqrt{t}$ which agrees with intermediate model \cite{2-n}, where $\phi\propto\sqrt{(1-f)t}$. The form of potential in logamediate model is $V(\phi)\propto\frac{(\ln\phi)^{\lambda-1}}{\phi^2}$ but in intermediate model tachyon potential has the form $V(\phi)\propto\phi^{-4\frac{f-1}{2f-1}}$. These two forms of potential have the special properties of the tachyon potentials. On the other hand the Harrison-Zeldovich spectrum of density perturbation (i.e. $n_s=1$) was obtained in warm-tachyon-intermediate inflation model \cite{2-n}, for fixed value of parameter $f$ where $f=\frac{2}{3}$, but in warm-tachyon-logamediate inflation model the scale invariant spectrum is not obtained for one fixed value of parameter $\lambda$.
In logamediate scenario, we could approximately obtain the Harrison-Zeldovich spectrum for large values of parameter $\lambda$ ($\lambda\geq50$).  FIGs.(1) and (3) show  the scale invariant spectrum may be presented for ($\lambda, N)=(50,60)$. FIG.(2)  shows the Harrison-Zeldovich spectrum could be approximately obtained for $(\lambda,\Gamma_1)=(50,37.5)$. FIG.(4) shows  the Harrison-Zeldovich spectrum could be approximately obtained for $(\lambda,\Gamma_0)=(50,37.5)$. Scale-invariant spectrum in logamediate scenario is given by two parameters ($\lambda,N$) or ($\lambda,\Gamma_i$), but in intermediate inflation the scale-invariant spectrum was presented by one parameter $f$.  \cite{2-n}
We also have found the small level of non-Gaussianity for large value of parameter $\lambda$ at the late time.


\end{document}